\documentclass[aapm,mph,reprint,onecolumn]{revtex4-1}
\usepackage{setspace}
\usepackage[mathlines]{lineno}% Enable numbering of text and display math
\usepackage[cmex10]{amsmath}
\usepackage{tabularx}
\usepackage{multirow}
\usepackage{booktabs}
\usepackage{color}
\usepackage{tikz}
\usepackage{array}

\usepackage{array}

\usepackage{changepage}

   \DeclareGraphicsExtensions{.pdf,.jpeg,.png,.eps}

\usepackage[normalem]{ulem}
\newcommand{\icomment}[1]{\textcolor{black}{#1}}

\newcommand{\ihighlight}[1]{\textcolor{black}{#1}}

\def\paragraph{\subsubsection*}

\newcommand{\beq}{\begin{equation}}
\def\eeq{\end{equation}}
\newcommand{\be}{\begin{enumerate}}
\newcommand{\ee}{\end{enumerate}}
\newcommand{\bi}{\begin{itemize}}
\newcommand{\ei}{\end{itemize}}
\newcommand{\bc}{\begin{center}}
\newcommand{\ec}{\end{center}}

\DeclareMathOperator*{\argmin}{arg\,min}

\begin{document}

% Use the \preprint command to place your local institutional report number
% on the title page in preprint mode.
% Multiple \preprint commands are allowed.
%\preprint{}

\title{\ihighlight{A unified material decomposition framework for quantitative dual- and triple-energy CT imaging}} %Title of paper

% repeat the \author .. \affiliation  etc. as needed
% \email, \thanks, \homepage, \altaffiliation all apply to the current author.
% Explanatory text should go in the []'s,
% actual e-mail address or url should go in the {}'s for \email and \homepage.
% Please use the appropriate macro for the type of information

% \affiliation command applies to all authors since the last \affiliation command.
% The \affiliation command should follow the other information.

\author{Wei~Zhao$^{1,2}$,~Don Vernekohl$^{1}$,~Fei~Han$^{1,3}$,~Bin~Han$^{1}$,~Hao~Peng$^{1}$,~Yong~Yang$^{1}$,~Lei~Xing$^{1*}$,~James K~Min$^4$}%
\email[]{lei@stanford.edu,jkm2001@med.cornell.edu}
%\homepage[]{Your web page}
%\thanks{}
%\altaffiliation{$^{2}$ Sun Yat-sen University Cancer Center, Department of Radiation Oncology, , Guangzhou, Guangdong, 94305 China.}
\affiliation{$^{1}$ Stanford University, Department of Radiation Oncology, Stanford, CA 94305 USA.}
\affiliation{$^{2}$ Department of Biomedical Engineering, Huazhong University of Science and Technology, Hubei, China.}
\affiliation{$^{3}$ Sun Yat-sen University Cancer Center, Department of Radiation Oncology, , Guangzhou, Guangdong, China.}

\affiliation{$^{4}$ Dalio Institute of Cardiovascular Imaging New York-Presbyterian Hospital and Weill Cornell Medical College, New York, NY 10021 USA.}

% Collaboration name, if desired (requires use of superscriptaddress option in \documentclass).
% \noaffiliation is required (may also be used with the \author command).
%\collaboration{}
%\noaffiliation

%\date{\today}

\begin{abstract}
% insert abstract here

\textbf{Purpose:} Many clinical applications depend critically on the accurate differentiation and classification of different types of materials in patient anatomy. This work introduces a unified framework for accurate nonlinear material decomposition and applies it, for the first time, in the concept of triple-energy CT (TECT) for enhanced material differentiation and classification as well as dual-energy CT.\\
\textbf{Methods:} \ihighlight{We express polychromatic projection into a linear combination of line integrals of material-selective images. The material decomposition is then turned into a problem of minimizing the least-squares difference between measured and estimated CT projections. The optimization problem is solved iteratively by updating the line integrals. The proposed technique is evaluated by using several numerical phantom measurements under different scanning protocols.} The triple-energy data acquisition is implemented at the scales of micro-CT and clinical CT imaging with commercial "TwinBeam" dual-source DECT configuration and a fast kV switching DECT configuration. Material decomposition and quantitative comparison with a photon counting detector and with the presence of a bow-tie filter are also performed.\\
\textbf{Results:} The proposed method provides quantitative material- and energy-selective images examining realistic configurations for both dual- and triple-energy CT measurements. Compared to the polychromatic kV CT images, virtual monochromatic images show superior image quality. For the mouse phantom, quantitative measurements show that the differences between gadodiamide and iodine concentrations obtained using TECT and idealized photon counting CT (PCCT) are smaller than 8 mg/mL and 1 mg/mL, respectively. \ihighlight{TECT outperforms DECT for multi-contrast CT imaging and is robust with respect to spectrum estimation}. For the thorax phantom, the differences between the concentrations of the contrast map and the corresponding true reference values are smaller than 7 mg/mL for all of the realistic configurations.\\
\textbf{Conclusions:} A unified framework for both dual- and triple-energy CT imaging has been established for the accurate extraction of material compositions using currently available commercial DECT configurations. The novel technique is promising to provide an urgently needed solution for several CT-based \ihighlight{diagnostic} and therapy applications, \ihighlight{especially for the diagnosis of cardiovascular and abdominal diseases where multi-contrast imaging is involved}.
\end{abstract}

% insert suggested PACS numbers in braces on next line

\maketitle %\maketitle must follow title, authors, abstract and \pacs

% Body of paper goes here. Use proper sectioning commands.
% References should be done using the \cite, \ref, and \label commands
\section{Introduction}
% The very first letter is a 2 line initial drop letter followed
% by the rest of the first word in caps.
%
% form to use if the first word consists of a single letter:
% \IEEEPARstart{A}{demo} file is ....
%
% form to use if you need the single drop letter followed by
% normal text (unknown if ever used by the IEEE):
% \IEEEPARstart{A}{}demo file is ....
%
% Some journals put the first two words in caps:
% \IEEEPARstart{T}{his demo} file is ....
%
% Here we have the typical use of a "T" for an initial drop letter
% and "HIS" in caps to complete the first word.
\label{sec:intro}

In computed tomography (CT) imaging, the CT number corresponds to the effective linear attenuation coefficient of a voxel and its value depends on the X-ray photon energy. In reality, it is possible that two voxels of the same material have different CT numbers (or effective linear attenuation coefficients), as the effective energy may be location dependent due to the beam hardening effect. Furthermore, the CT number of a voxel depends on both the atomic number and the mass density of the material, making material decomposition from a single-energy CT impossible without additional information. To circumvent the problem, \ihighlight{dual-energy CT (DECT) has been developed to allow for material discrimination, by adding another attenuation measurement with a different energy spectrum}~\cite{alvarez1976,johnson2007}. DECT has been a valuable tool in a range of emerging clinical applications, including automated bone removal in angiography, blood volume measurement in perfusion CT, urinary stone characterization and gout diagnosis~\cite{mccollough2015}. Practically, dual-energy attenuation measurements have been implemented in different ways, including the fast switching of X-ray tube potentials in a single scan ~\cite{kalender1986,silva2011}, dual X-ray sources on the same gantry~\cite{johnson2007}, and layered detectors~\cite{carmi2005}. The measured dual-energy projection data sets are then employed to yield energy- and material-selective images using dual-energy material decomposition methods. Depending on the data acquisition technique, dual-energy material decomposition can be performed in projection domain, image domain, or \ihighlight{projection- and image- domain jointly}.

	Like many other material decomposition methods~\cite{Yan2000,Sullivan2004,sidky2004,stenner2007,noh2009,brendel2009,chen2016,ducros2017,hohweiller2017,pham2017}, the linear attenuation coefficient is expressed as a linear combination of two basis materials in projection domain methods. \ihighlight{Thus, the polychromatic high- and low-energy projections become functions of the line integrals of the basis material images. This reduces the problem of material decomposition to the determination of the integrals of these basis material images using the high- and low-energy projections and a decomposition function. The decomposition function can be approximated by, for example, a polynomial where its coefficients are determined by calibrations with dedicated experimental phantoms~\cite{stenner2007}}. Basis material sinograms obtained using the decomposition functions are then employed to reconstruct the basis material images using standard image reconstruction algorithms. Alternatively, dual-energy material decomposition can be implemented in image-domain~\cite{niu2014,clark2014}. Here, high- and low-energy CT images are first reconstructed using the corresponding raw projections and material decomposition is then performed in image domain by linearly combining the CT images. As the acquired high- and low-energy CT data are usually geometrically inconsistent in practice (i.e., the paths of the two measurements are different, even for the fast kVp switching technology), the material decomposition in image-domain is more convenient. \ihighlight{Like projection-domain methods, a series of calibration measurements for the basis materials in high- and low-energy are also required. However,} material-selective images generated this way are basically a linear combination of the high- and low-energy CT images. Compared to nonlinear projection domain methods, this leads to reduced contrast-to-noise ratio and residual artifacts in the decomposed images~\cite{kuchenbecker2015}.

Dual-energy material decomposition in the third way generates the material images directly from dual-energy attenuation measurements with incorporation of a projection matrix~\cite{sukovic2000,zhang2014,long2014,barber2016}. In this case, material decomposition is formulated as an optimization problem, and statistical models and regularization are often introduced into the objective function to reduce noise and improve image quality. Material decomposition in this approach is usually computationally expensive because of repeated forward and backward projections as well as complicated optimization procedures, which may be undesirable for clinical applications.

Although DECT material decomposition yields images with significantly reduced beam hardening artifacts and allows a quantitative characterization of two different materials (e.g., contrast medium and calcified tissues) ~\cite{tran2009}, there are many applications in which the discrimination of more than two materials \ihighlight{is} needed. Indeed, the human body contains different materials, such as blood, water, bone, and often endogenous contrast agent during imaging. Thus a material decomposition method capable of discriminating more than two materials is highly desirable. In general, the problem of obtaining more than two material-specific images can be solved either in image-domain~\cite{liu2009,mendoncca2014,xue2017} or with energy-resolving photon counting detectors~\cite{roessl2007,bornefalk2010,modgil2015,zimmerman2015,zhang2017}. The former approach, however, is not as accurate as projection-domain methods because of residual beam hardening artifacts. The use of photon-counting detectors is still at an early stage and there are a number of technical issues, such as charge sharing~\cite{lee2016} and pulse pileup, yielding to distorted spectrum information, which have \ihighlight{yet} to be resolved. While it is possible to use more than two energy measurements for multi-material decomposition~\cite{granton2008,yu2016}, currently, only methods in image-domain exist to accomplish the task. The purpose of this work is to establish a unified nonlinear projection-domain framework for both dual- and triple-energy material decomposition.% with an accurate consideration of the X-ray energy spectrum.

%\hfill mds
%
%\hfill August 26, 2015

\section{Methods}
\subsection{Principle of basis material decomposition}
For diagnostic X-ray imaging, the attenuation properties of materials can be modeled using the two dominant physical interactions, i.e., photoelectric and Compton. Based on this assumption, in the theory of material decomposition, the linear attenuation coefficient $\mu(\vec{r},E)$ of a material is assumed to be modeled as a linear combination of two basis functions,
\beq\label{equ:decompdual}
\mu(\vec{r},E)=f_{1}(\vec{r})\psi_{1}(E)+f_{2}(\vec{r})\psi_{2}(E),
\eeq
here \ihighlight{$\vec{r}$ represents the three-dimensional spatial coordinates of a specific pixel and $E$ is the photon energy}. Basis functions $\psi_{1,2}$ are the independent energy dependencies which can be mass attenuation coefficients of known materials (i.e., basis materials) and $f_{1,2}(\vec{r})$ are material-specific images (specific to the basis materials).
However, if an element has a K-edge in the diagnostic energy range (for example, gadolinium), then the attenuation coefficient of the material cannot be decomposed solely with two basis materials when bone is also present. \ihighlight{This is because the specific K-edge can not be characterized by either Compton or photonelectric which are the dominant absorption mechanisms in the diagnostic energy range.} Hence, the decomposition in~(\ref{equ:decompdual}) has to be extended to include a third term, which can characterize the attenuation properties of the element. In this case, the decomposition of the material becomes,
\beq\label{equ:decomptriple}
\begin{split}
\mu(\vec{r},E)&=f_{1}(\vec{r})\psi_{1}(E)+f_{2}(\vec{r})\psi_{2}(E)+f_{3}(\vec{r})\psi_{3}(E)\\
&=\sum_{i=1}^3 f_{i}(\vec{r})  \psi_{i}(E)
\end{split}
\eeq
where $\psi_{i}(E)$ with $i=1,2,3$ are the known mass attenuation coefficients of the basis materials, including the K-edge material, and $f_{i}(\vec{r})$ are the location-dependent material-specific images. Without extra conditions, the extended formulation~(\ref{equ:decomptriple}) implies the need of a third attenuation measurements to obtain material-specific information.
% needed in second column of first page if using \IEEEpubid
%\IEEEpubidadjcol

\subsection{Material images-based polychromatic reprojection}

Based on the material decomposition formulation~(\ref{equ:decompdual}) and~(\ref{equ:decomptriple}), the polychromatic X-ray transmission measurement of the material can be estimated as
\beq\label{equ:polyreprojBimg}
\hat{p}=-\mathrm{log}\int_{0}^{E_{max}}\mathrm{d}E\,\Omega(E) \, \eta(E)\,\mathrm{exp}\left[-\sum_{i=1}^N \psi_{i}(E) A_i\right],
\eeq%-\sum_{j=1,2}A_{j}\psi_{j}(E) f_{i}(\vec{r})
with $N=2$ for two material case and $N=3$ for three material case, and with $A_{i}=\int\mathrm{d}\vec{r}\,f_{i}(\vec{r})$ the line integrals of the material-selective images (i.e. material density images). $\Omega(E)$ is the corresponding polychromatic energy spectrum of the ray and the integral in~(\ref{equ:polyreprojBimg}) goes over the energy range of the spectrum. $\eta(E)$ is the energy dependent detection efficiency. Note that $\hat{p}$ depends on the detector pixel and the pixel channel index is dropped for convenience.

\begin{figure*}[htbp]
    \centering
    \includegraphics[width=6.0in]{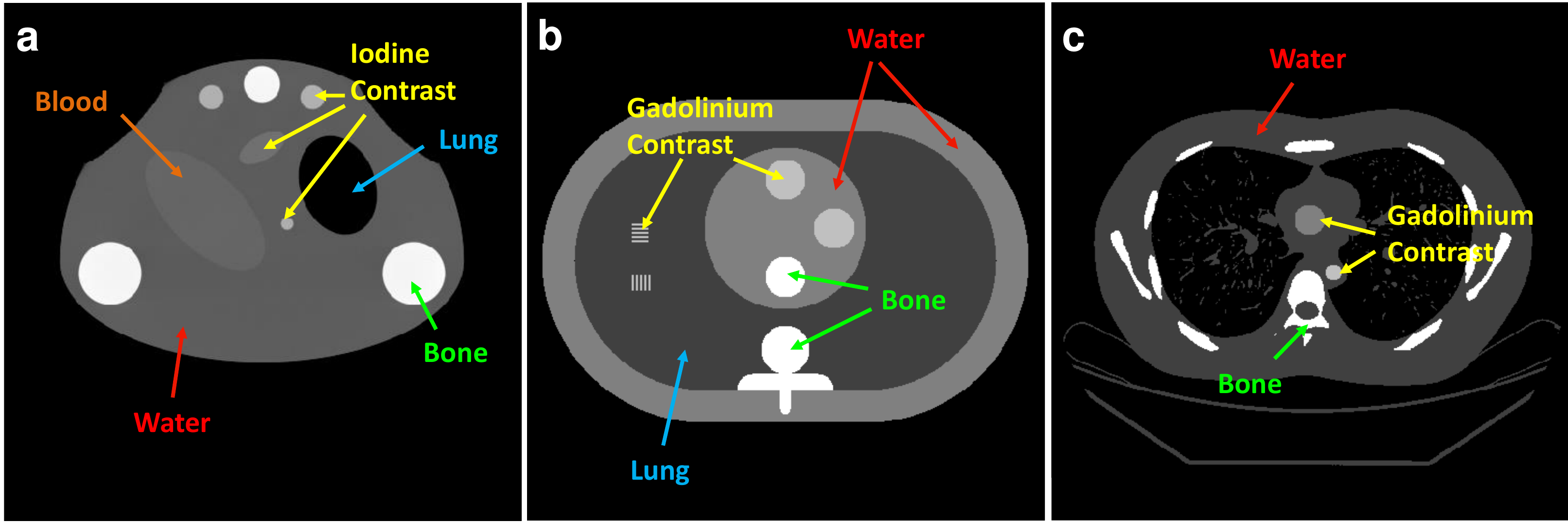}
    \caption{Imaging phantoms used in the numerical studies. (a) Modified micro-CT mouse phantom with blood, lung, bone and contrast agents inserts. The iodine contrast agents have two concentrations, 5 mg/mL and 10 mg/mL. (b) Modified thorax phantom showing two coronary arteries in the heart region and two bar patterns in the lung region both filled with gadolinium contrast agents. (c) Anthropomorphic thorax modified from clinical CT image showing contrast agents in the heart region.}%The bone inserts were 200 mg/mL hydroxylapatite (HA), Ca$_5$(PO$_4$)$_3$OH.
    \label{fig:f1}
\end{figure*}

\subsection{Material decomposition}

As mentioned above, to enable a three material decomposition, i.e., calculate the line integrals of the three material-specific images, one has to perform three transmission measurements using different energies, otherwise the system is underdetermined. Meanwhile, based on the polychromatic reprojection~(\ref{equ:polyreprojBimg}), for each detector channel, the reprojected value $\hat{p}$ should equal to the corresponding raw measurement $p_\mathrm{m}$. However, in realistic application, the acquired energy spectrum $\Omega(E)$ is prone to suffer from error, additionally, the residual scatter radiation as well as the projection noise (quantum noise and electronic noise) cause inconsistency between the measured data and the data calculated using reprojection model. Hence, instead of directly solving the line integrals of the material-specific images from~(\ref{equ:polyreprojBimg}), we minimize the quadratic error between the measured data $p_\mathrm{m}$ and the estimated projection $\hat{p}$ with respect to the line integrals $A_i$, i.e.,
%\beq\label{equ:calibration}
%(A_1^{\ast},A_2^{\ast},A_3^{\ast})=\argmin_{A_1,A_2,A_3\geq0}\|\textbf{p}_m-\hat{\textbf{p}}(A_1,A_2,A_3)\|_2^2.
%\eeq%-\sum_{j=1,2}A_{j}\psi_{j}(E) f_{i}(\vec{r})

\beq\label{equ:calibration}
A_i^{\ast}=\argmin_{A_i\geq0}\|\textbf{p}_\mathrm{m}-\hat{\textbf{p}}(A_i)\|_2^2.
\eeq%-\sum_{j=1,2}A_{j}\psi_{j}(E) f_{i}(\vec{r})

Here $i=1,2$ for dual-energy case, and $i=1,2,3$ for triple-energy case. The vector $\textbf{p}_\mathrm{m}$ denotes the dual- or triple-energy transmission measurements for each detector pixel, i.e., $\textbf{p}_\mathrm{m} = (p_\mathrm{m}^H,p_\mathrm{m}^L)$ or $\textbf{p}_\mathrm{m} = (p_\mathrm{m}^H,p_\mathrm{m}^M,p_\mathrm{m}^L)$, where $p_\mathrm{m}^H$, $p_\mathrm{m}^M$, and $p_\mathrm{m}^L$ are the high-, median-, and low-energy measurements, respectively. $\hat{\textbf{p}}$ is the corresponding estimated projection using line integrals of the density images. Since $A_i$ is the line integral of the density image, we have non-negative constraint on $A_i$. The optimization problem (\ref{equ:calibration}) can be solved in a pixel-wise fashion using a multi-variable downhill simplex method (Nelder-Mead method). \ihighlight{Implementation details and pseudo code of the method can be found in \ref{sec:appen}.} Once $A_i$ was determined for all of the pixels, material-specific density images can be reconstructed by using the standard filtered backprojection algorithm.

\section{Simulation studies}

In order to evaluate the proposed method, we performed simulation studies using various phantoms in scale of both clinical CT and micro-CT applications.

\subsection{Numerical phantoms description}

We simulated three test phantoms to show the performance of the nonlinear material decomposition method. The first mouse phantom which is in scale of micro-CT application and modified from the phantom as defined by Stenner and Kachelriess in~\cite{stenner2007}, consists of inserts that contain different iodine concentrations. Compared to the original phantom in~\cite{stenner2007}, we have replaced the low contrast and the bone inserts with blood and standard cortical bone inserts, respectively. Concentrations for the two iodine contrast agent inserts are 5 and 10 mg/mL. %The material used for the bone inserts is the 200 mg/mL bone-mineral hydroxylapatite (HA), Ca$_5$(PO$_4$)$_3$OH, and has a density of 1.16 g/cm$^3$. Due to its low density, the bone inserts would have similar attenuation with the contrast agents, which is difficult to differentiate from each other using conventional CT techniques.

The second phantom is a simplified thorax phantom and is in scale of clinical diagnostic CT application. It contains two coronary arteries filled with contrast agent in the heart region. In order to depict the preservation of spatial resolution, two contrast agent bar patterns are also included in the lung region. Compared to the second one, the third anthropomorphic thorax phantom is more realistic as it was generated using clinical CT images. To perform triple-energy material decomposition, we have added two contrast agent inserts in the heart region. Since both of the thorax phantoms are scanned using clinical CT protocols which have relatively high kV settings and thick inherent filtrations, we have used gadolinium (Gd) as contrast agent for both thorax phantoms as the K-edge energy for gadolinium is 50.2 keV while it is 33.2 keV for iodine. Specifically, we model OMNISCAN$^{\textbf{TM}}$ as the contrast agent which contains 287 mg/mL gadodiamide and with a mass density of 1.15 g/cm$^3$. The gadodiamide has the stoichiometric formula of C$_{16}$H$_{28}$GdN$_5$O$_9$. To be more realistic, concentration of gadodiamide of the OMNISCAN$^{\textbf{TM}}$ is diluted to 60 mg/mL. The phantoms used in this study are shown in Fig.~\ref{fig:f1}.

\begin{figure}[t]
    \centering
    \includegraphics[width=2.8in]{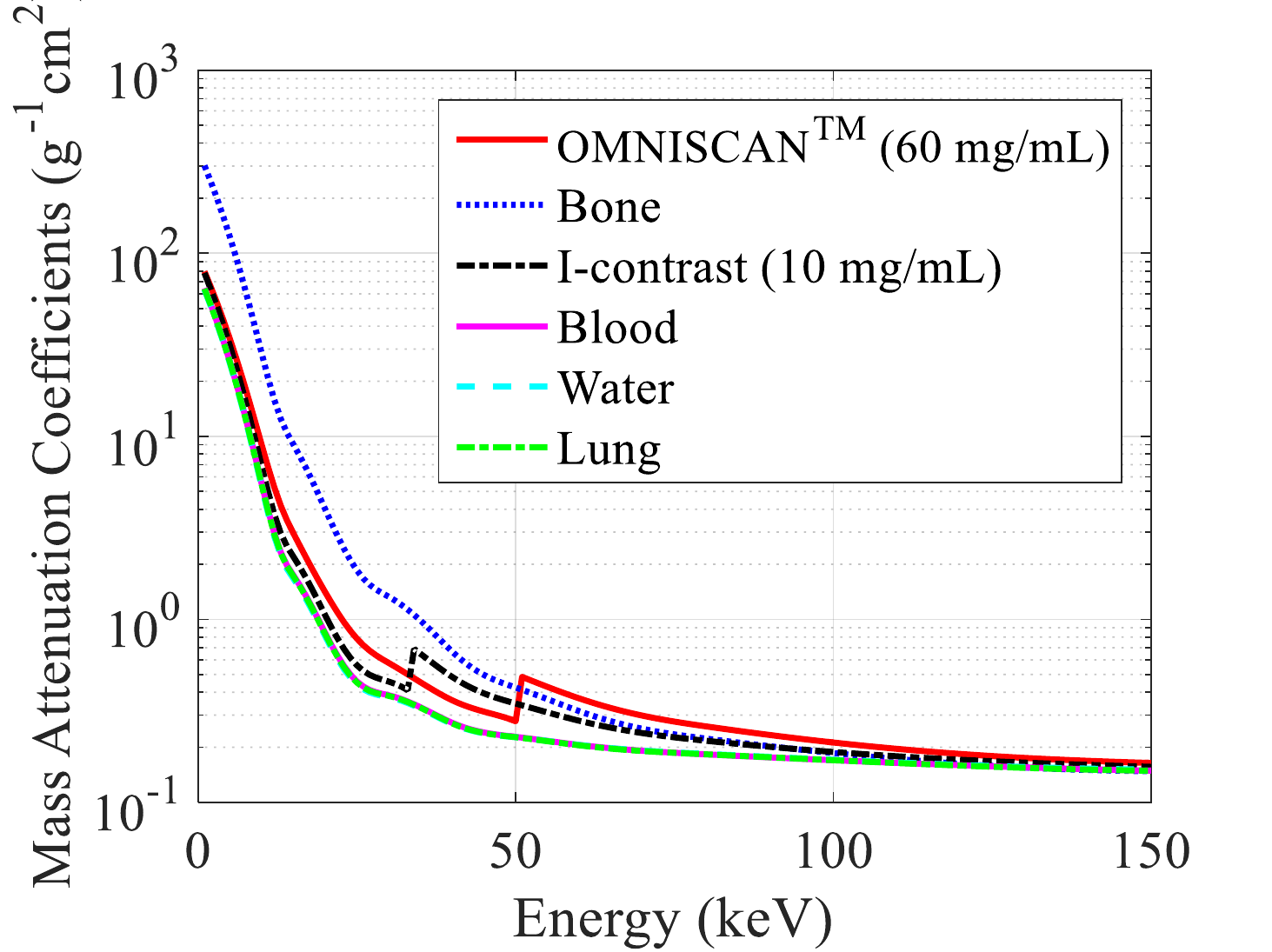}
    \caption{Mass attenuation coefficients of the materials in this study. Note that although blood, water, and lung have similar mass attenuation coefficients, their linear attenuation coefficients are different due to their different mass density.}
    \label{fig:f2}
\end{figure}

During the simulation, all of the mass attenuation coefficients are obtained or calculated from the NIST database. For the materials which can not be found in the database, such as the contrast agents, their mass attenuation coefficients are calculated using their chemical elements and the corresponding mass fractions. Figure~\ref{fig:f2} shows the mass attenuation coefficients of the materials used in the simulation studies. Note that the y axis is plotted on a logarithmic scale. One can observe the K-edge energies of the iodine and gadolinium contrast agents. Properties of the materials used in this study are summarized in Table~\ref{tab:materials}.

\begin{table}[h]
\centering
\caption{Properties of the materials used in the numerical simulation studies. Note that we have diluted the OMNISCAN$^{\textbf{TM}}$ to 60 mg/mL and \icomment{its} density is reduced from 1.15 g/cm$^3$ to 1.03 g/cm$^3$.}
\label{tab:materials}
\begin{center}
\begin{tabular}{ll} %% this creates two columns
%% |l|l| to left justify each column entry
%% |c|c| to center each column entry
%% use of \rule[]{}{} below opens up each row
\toprule
\rule[-1ex]{0pt}{3.5ex}  Material name & Density (g/cm$^3$) \\ %\hline{2-3}\multicolumn{2}{c}{Value}
\hline
\rule[-1ex]{0pt}{3.5ex}  Water & $1.000$  \\
\rule[-1ex]{0pt}{3.5ex}  Lung & $0.260$   \\
\rule[-1ex]{0pt}{3.5ex}  Bone & $1.850$  \\
\rule[-1ex]{0pt}{3.5ex}  Blood & $1.060$  \\
\rule[-1ex]{0pt}{3.5ex}  Iodine contrast agent (5 mg/mL, 10 mg/mL) & $1.004, 1.008$   \\
%\rule[-1ex]{0pt}{3.5ex}  Gadolinium contrast agent (30 mg/mL) & $1.026$  \\
\rule[-1ex]{0pt}{3.5ex}  OMNISCAN$^{\textbf{TM}}$ (60 mg/mL) & $1.031$  \\ % because the undiluted is low
\bottomrule
\end{tabular}
\end{center}
\vspace{-1em}
\end{table}

%\subsection{Simulated materials}
%
%Based on~(\ref{equ:calibration}), material decomposition is performed using raw projection data sets. For dual-energy case, if K-edge material is present in the phantom, one needs a third basis material to characterize the attenuation of the K-edge material, so we perform dual-energy material decomposition using the micro-CT mouse phantom without the presence of iodine contrast agent. In this case, water and bone are chosen as the basis materials. To evaluate triple-energy material decomposition, all of the three phantoms are employed and water, bone, as well as the corresponding contrast agent (i.e., iodine for the mouse phantom and gadolinium for the thorax phantoms) are used for the basis materials during decomposition.

\subsection{Energy spectra}

For the micro-CT mouse phantom study, we use 40 kV, 60 kV and 80 kV spectra for the triple-energy transmission measurements. \ihighlight{These three spectra are generated using the most recent spectrum generator, Spektr 3.0}~\cite{punnoose2016}, with 0 mm Al, 5 mm Al, and 0.3 mm Cu filtrations, respectively. In addition, all of the three sources have 2 mm Al inherent filters. We use the low 40 kV and the high 80 kV spectra for dual-energy measurements and all the three spectra for triple-energy CT measurements. \ihighlight{In order to accurately calculate $\hat{p}$, the energy spectrum used in~(\ref{equ:polyreprojBimg}) should be modeled precisely. There are a lot of methods that can be employed to obtain the spectrum~\cite{zhao2015itm,duisterwinkel2015,chang2016,leinweber2017x,zhao2017JMI}. In this study, we employ an indirect transmission measurement (ITM) method to estimate the triple energy spectra~\cite{zhao2015itm}. Specifically, the energy spectrum is expressed by a set of model spectra. By updating each weight of the model spectra, the spectrum estimation method minimizes the difference between the ITM and the projection of a uniform water phantom. The final spectrum is calculated using the model spectra and their weights. The estimated spectra are then employed for the triple-energy material decomposition. To further evaluate the robustness of the proposed material decomposition method over spectrum estimation accuracy, triple-energy material decomposition is performed using spectra estimated with two sets of different model spectra. }

For the triple-energy thorax phantoms studies, it is natural to take advantage of the current commercial dual-energy CT (DECT) configurations to acquire triple-energy CT data sets as DECT scanners are widely used in clinical applications nowadays. This is beneficial over sequentially scanning the subject with three different energies or using three source-detector sets on the same gantry. For the dual-source DECT scanner, one can use a "TwinBeam" configuration on one of the X-ray sources~\cite{yu2016} to enable the simultaneous acquisition of a triple-energy measurement. With the "TwinBeam" configuration, X-ray photons emitted from the source tube are prefiltered before arriving at the subject by two different filters, and each filtered spectrum covers half of the detector rows.

To reduce noise during material decomposition, one would expect the filtered spectra should be quite different from each other such that the difference between their transmission measurements of the sample is as large as possible. To this end, it is better to choose K-edge filters to separate the spectra. Because K-edge filters significantly attenuate X-ray photons whose energy are higher than the K-edge energy. This feature can be exploited to generate high- and low-energy spectra, where the high-energy spectrum is generated by the filter with low K-edge energy and the low-energy spectrum is generated by the filter with high K-edge energy. For example, as shown in~\cite{yu2016}, for a 150 kV incident spectrum, tin (Sn) filter is employed to generate a hardened spectrum, while gold (Au) and bismuth (Bi) filter set is used to soften the spectrum.

\begin{figure}[t]
    \centering
    \includegraphics[width=3.5in]{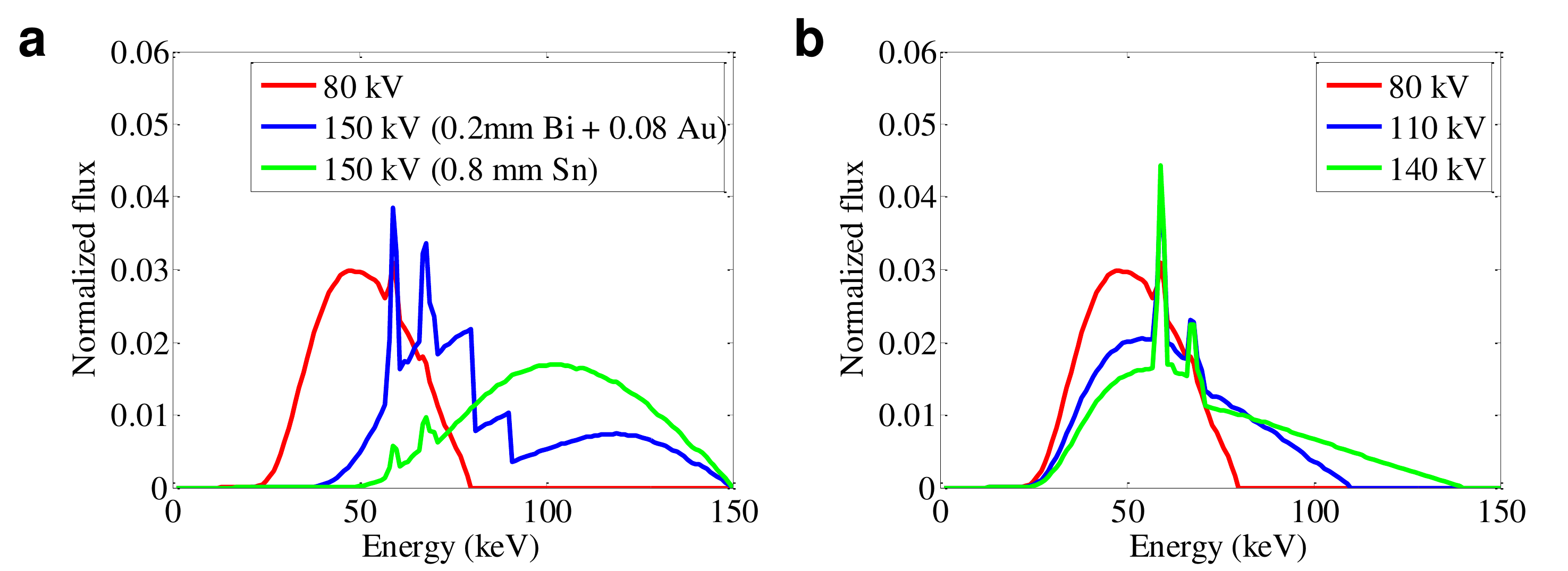}
    \caption{Energy spectra used in the numerical simulation studies of clinical diagnostic CT applications. (a) Siemens scenario, Bi and Au filters are used to filter high energy photons to yield a median energy spectrum, while Sn filter is used to filter low energy photons to yield a high energy spectrum. (b) GE scenario, fast kV switching spectra with the same filtration.}
    \label{fig:f3}
\end{figure}

In this study, to evaluate the proposed method, we performed material decomposition using two different sets of spectra, "TwinBeam" dual-source spectra and fast kV switching spectra, i.e., Siemens scenario and GE scenario. For the Siemens scenario, a 0.6 mm Sn was used to filter a 150 kV spectrum to yield high-energy spectrum and 0.08 mm Au + 0.1 mm Bi recombination filters were used to yield median-energy spectrum. These two spectra together with a 70 kV spectrum make up the triple-energy CT configuration.
For the GE scenario, instead of fast switching between high kV and low kV, the scanner performs an additional median kV measurement for triple-energy CT. Since it is technically difficult to change filtration during CT scan in this configuration, all of the triple-energy measurements use the same filtration. Figure~\ref{fig:f3} depicts the spectra used for the simulation studies. For the "TwinBeam" dual-source DECT configuration (Fig.~\ref{fig:f3} (a)), one can achieve triple-energy measurement using different kVs as well as different filtrations. Since the K-edge energies for bismuth (Bi) and gold (Au) are 80.7 and 90.5 keV, respectively, the bismuth and gold filters are used to filter high energy photons to yield a median-energy spectrum for the triple-energy measurement. To the contrary, tin filter whose K-edge energy is 29.2 keV, is used to filter low-energy photons and yield high-energy spectrum. For the fast kV switching configuration (Fig.~\ref{fig:f3} (b)), the triple-energy measurements are performed using different kVs but the same filtration.

\begin{figure}[t]
    \centering
    \includegraphics[width=3.5in]{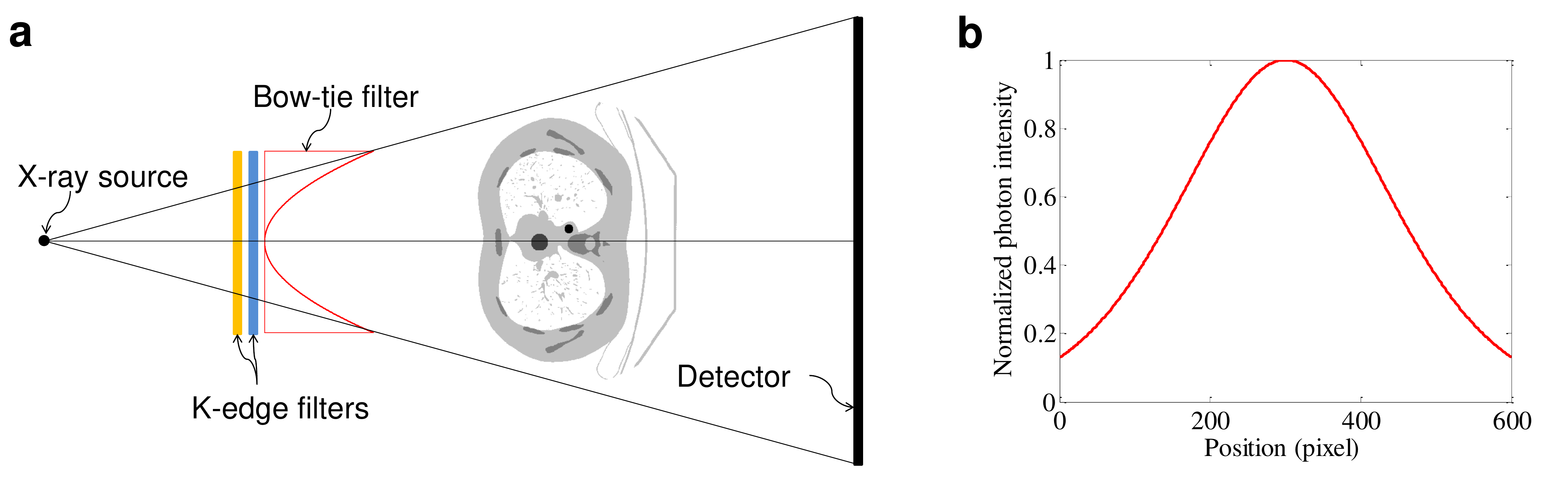}
    \caption{Anthropomorphic thorax phantom study using bow-tie filter. (a) CT scan geometry with the bow-tie filter and K-edge filters. (b) Flood field of the bow-tie filter.}
    \label{fig:f4}
\end{figure}

To be more realistic, the proposed method was also tested on the anthropomorphic thorax phantom including a bow-tie filter in the acquisition geometry. In this case, the Siemens DECT configuration was used with the bow-tie filter attached behind the K-edge filters, i.e., between the K-edge filters and the phantom. Figure~\ref{fig:f4}(a) shows one source-detector set for this configuration. Figure~\ref{fig:f4}(b) shows the normalized photon intensity of the detector without the presence of the phantom. All the used spectra for the thorax phantoms studies are generated using Spektr 3.0~\cite{punnoose2016}.

\subsection{Projection data acquisition}

All of the simulations were performed using 2D fan-beam geometry \ihighlight{with an ideal point X-ray source. The linear detector array consists of 1024 pixels and each pixel is 0.388 mm$\times$ 0.388 mm}. Since one difficulty of material decomposition is ill-conditioning, \ihighlight{which yields material images with increased noise}, Poisson noise is included in the simulations to validate the robustness of the projection domain material decomposition method. The fan beam CT projection data sets were simulated by polychromatic forward projecting the micro-CT phantom and the thorax phantoms. Mathematically, the projection data can be represented as:

\beq\label{equ:weight}
\begin{split}
I &= \\
&\int_{0}^{E_{max}}\mathrm{d}E\, \eta(E)\,\mathrm{Poisson} \left\{N\Omega(E) \,\mathrm{exp}\left[-\int\mu(\vec{r},E)\mathrm{d}\vec{r}\right]\right\},
\end{split}
\eeq
with $N$ the total number of incident photons and it was set to the order of $10^6$ for all of the clinical CT studies. To be realistic, we simulated an energy-integrating detector, thus the energy dependent response $\eta(E)$ was considered to be proportional to photon energy $E$. $E_{max}$ is the maximum energy of the polychromatic spectrum $\Omega(E)$. In the energy integrating mode, all the photons arriving on a specific detector pixel for a single view angle are summed up according to their energies. The energy-dependent linear attenuation coefficient $\mu(\vec{r},E)$ was obtained from the National Institute of Standards and Technology (NIST) database. During the simulations, the propagation path length for each ray in each material is calculated using either an analytical method or a numerical ray-tracing method. \ihighlight{To generate a Poisson distributed photon count, we first calculate the mean photon number $N'(E)=N\Omega(E) \,\mathrm{exp}\left[-\int\mu(\vec{r},E)\mathrm{d}\vec{r}\right]$ that arrive at the detector for each energy bin. A random number is then generated from the Poisson distribution with $N'(E)$  as the mean parameter~\cite{whiting2006,wang2009noise,zhang2014noise}. These random photon numbers are weighted by their energies (energy integration) and detection efficiencies and finally summed up.}

For the simple micro-CT mouse phantom, note that it consists of ellipses and circles which have explicit mathematical forms, we can use the analytical method. In this case, the propagation path length for each pixel in each ellipse or circle is the intersection length of the line (i.e., the ray connecting the pixel and X-ray source) and the specific ellipse or circle. Thus it can be analytically solved using the geometry equations of the line, ellipse and circle. For the complicated thorax phantoms, we have to use the numerical method. In this case, the intersection length of the ray and each voxel (with size of 0.45 mm) is numerically calculated and added up for the same material to yield the corresponding propagation path length. Due to its computation load and parallel feature, this method is implemented on a graphics processing unit (GPU)~\cite{pratx2011} for considerable acceleration.

%Based on the material images, the energy-selective monochromatic images can be calculated using equation~\ref{equ:decomposition}.
%
%For the bow-tie case study, the distance from the bow-tie filter to the x-ray source is
%The diameters of the water cylinder and the inserts were 35.2~mm and 4.8~mm, respectively. The second one is a oval phantom with water background and four iodine concentration inserts. For the clinical scenarios, numerical CT phantoms were studied. One is a  lung phantom consisting of cortical bone and soft tissue, and the other one is a thorax phantom consisting of cortical bone, soft tissue, adipose, and a bar pattern made of bone. For the micro-CT cylinder phantom, the simulation was carried out at 45 and 65 kVp. For clinical CT phantoms, the simulations were carried out at 100 and 140 kVp.
%
%Dual-energy material decomposition
%
%comparison study->multi-material decomposition using photon-counting detector

\subsection{Comparison studies}

Simultaneous multi-contrast CT imaging is very useful for the diagnosis of cardiovascular and abdominal diseases. The relevant clinical applications include characterization of atherosclerotic plaque composition~\cite{cormode2010}, colonography~\cite{muenzel2016}, and so on. In this case, DECT cannot provide accurate contrast images, while triple-energy CT (TECT) can provide quantitative multi-contrast images. In order to show the merit of TECT, a modified mouse phantom where the blood insert is replaced by gadolinium contrast and the bone inserts are replaced by air, is investigated using both DECT and TECT. \icomment{In addition, comparison study with a widely used sinogram domain method, EDEC method~\cite{stenner2007}, is also included. The EDEC method is an empirical nonlinear algorithm based on calibration measurements using basis materials. In this study, we have employed a two-cylinder phantom for calibration}.

Photon counting detector-based CT imaging is getting more and more attention from clinical side recently~\cite{atak2015,yuz2016}. It has several superior features where quantitative imaging and direct tissue characterization are the most important features in clinical applications. To further evaluate the performance of the proposed method, the method is compared to a nonlinear method based on photon counting detectors~\cite{roessl2007}. The latter method models the projection counting data as Poisson signals and formulates the material decomposition process as a maximum-likelihood optimization problem.

\section{Results} \label{sec:results}

We first show the results of dual-energy material decomposition and comparison study with \ihighlight{idealized} photon counting CT (PCCT) using the micro-CT mouse phantom, and then we present the TECT results using simplified and anthropomorphic thorax phantoms in the scale of clinical applications. Result with the application of bow-tie filter is also presented.

%\begin{figure}
%    \centering
%    \includegraphics[width=0.7\textwidth]{figure1-2.pdf}
%    \caption{Results of the numerical cylinder phantom. The 45 and 65 kV images, and the monochromatic image are windowed to (C=0 HU/W=500 HU), and the window setting for the water and iodine images is (C=100\%/W=40\%). Note that the basis images are windowed individually for a better visualization.}
%    \label{fig:f1}
%\end{figure}
%
%\begin{figure}[t]
%    \centering
%    \includegraphics[width=0.9\textwidth]{figure3_1.pdf}
%    \caption{Results of the numerical cylinder phantom using the proposed method at different number of photons. \emph{H} and \emph{L} stand for the number of high- and low-energy photons per ray, respectively. For all of the dose levels, beam hardening artifacts are perfectly removed.}
%    \label{fig:f3}
%\end{figure}
\begin{figure}[t]
    \centering
    \includegraphics[width=3.5in]{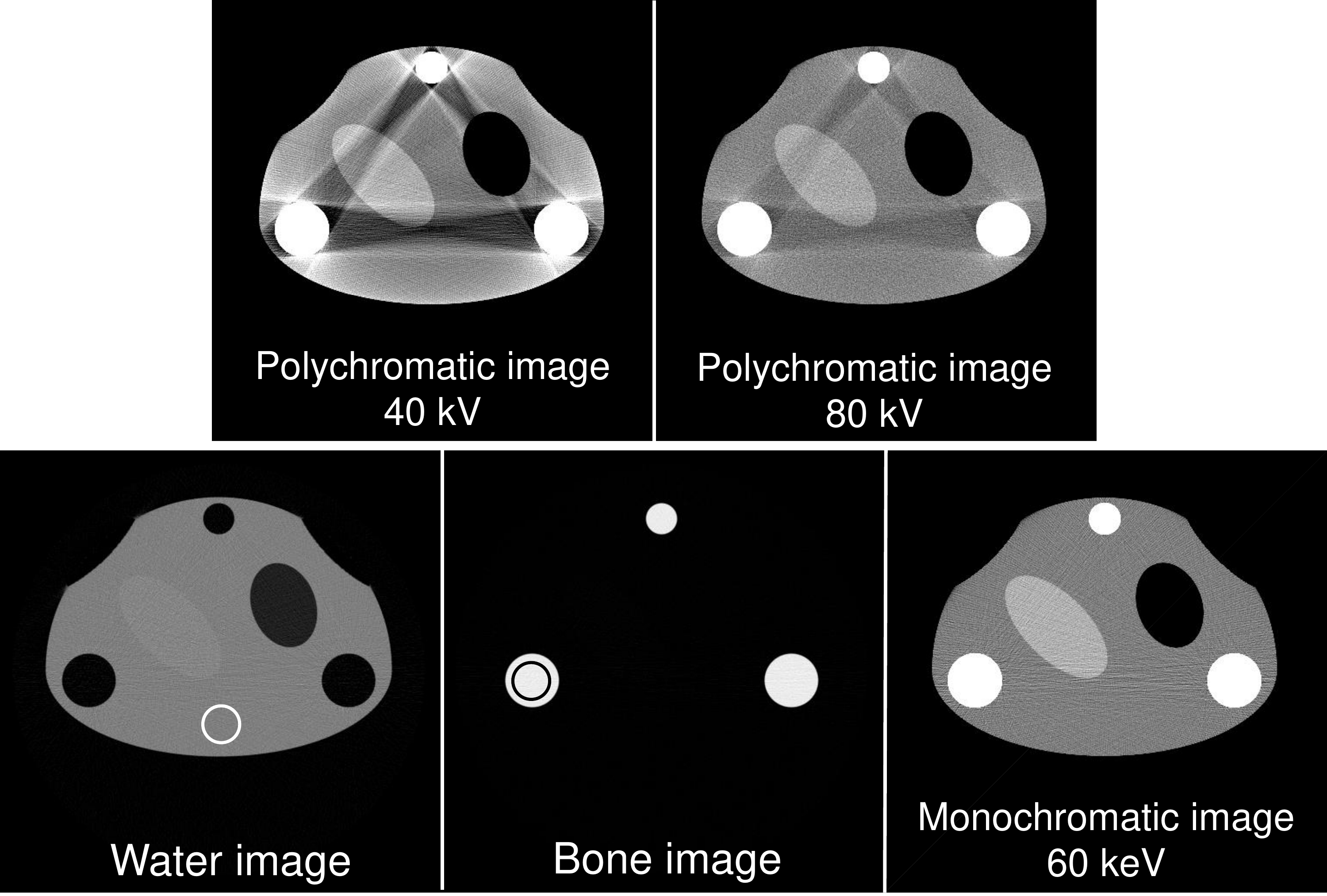}
    \caption{Dual-energy CT images of the micro-CT mouse phantom without contrast agent at 40 kV/80 kV, and material-selective images obtained using the proposed method. The beam-hardening artifacts free monochromatic image is a linear combination of the basis material images. Display window for the CT images and monochromatic image: [-200, 200] HU, and for the material images [0, 2] g/cm$^{3}$).}
    \label{fig:f5}
\end{figure}

\subsection{Dual-energy material decomposition}

Figure~\ref{fig:f5} shows the results of dual-energy material decomposition using the mouse phantom. The first row depicts polychromatic 40 kV and 80 kV CT images. It is visible that there are severe beam hardening artifacts in the CT images, especially in the 40 kV image. \ihighlight{This is because for the 40 kV spectrum, the linear attenuation coefficients of the materials undergo a larger change, compared to 80 kV spectrum.} The second row shows the basis material images and the monochromatic image at 60 keV. The monochromatic image is the weighted summation of the basis material images with the weights from linear attenuation coefficients of basis materials at 60 keV. Quantitative measurements of the decomposed material images match the true densities of blood and bone quite well. Compared to the standard polychromatic CT images, beam hardening artifacts are completely removed in the composed virtual 60 keV image. Note that the CT images and the monochromatic image are normalized to Hounsfield units (HU) with value measured in the central water region and with the linear attenuation coefficient of water at 60 keV, respectively.

\subsection{Triple-energy material decomposition}

\begin{figure}[t]
    \centering
    \includegraphics[width=3.5in]{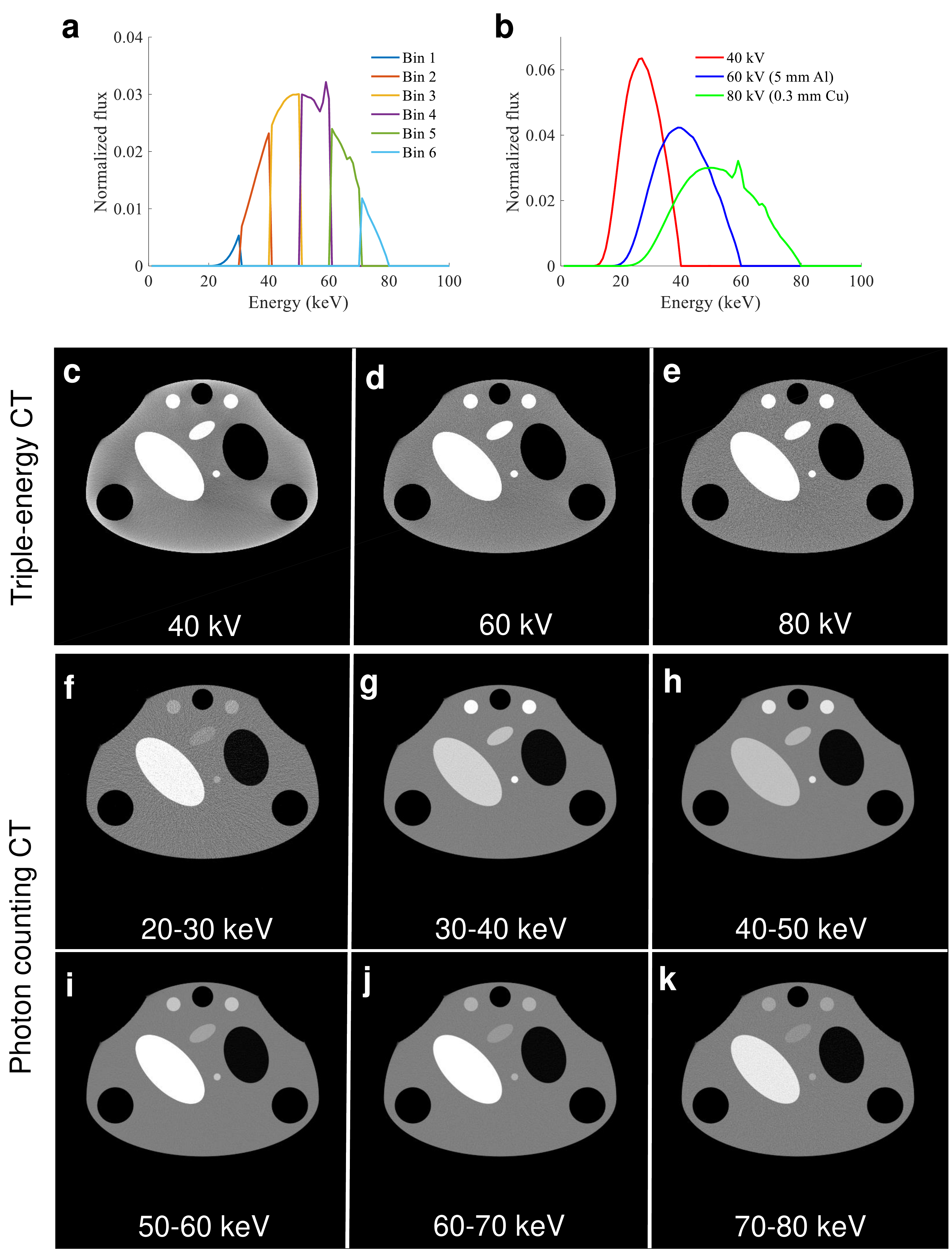}
    \caption{Results of CT images of the \ihighlight{mutil-contrast} micro-CT mouse phantom using both \ihighlight{idealized PCCT} and TECT. (a) Energy spectra for \ihighlight{ideally separated photon counting bins} of the PCCT, (b) energy spectra used for triple-energy measurements. (c)-(e) Polychromatic CT images reconstructed using projection acquired with TECT, and (f)-(k) images reconstructed using each energy bin of PCCT. All CT images are normalized to Hounsfield units and windowed to (C=0 HU/W=400 HU).}
    \label{fig:f6}
\end{figure}

%Comparison with PCCT-based material decomposition
The results of TECT imaging using the micro-CT mouse phantom are shown in Fig.~\ref{fig:f6}. For comparison, we also present the results of the phantom using idealized photon-counting detector. Figure~\ref{fig:f6} (a) and (b) show the spectra of \ihighlight{ideally separated photon counting bins} for PCCT and spectra for the triple-energy measurements, respectively. Different from dual-energy case, we have added a third 60 kV measurement between the low- and high-kV energy measurements. Figure~\ref{fig:f6}(c-e) depict the reconstructed polychromatic CT images using triple-energy measurements. Beam hardening artifacts are introduced in the 40 kV images, especially at the peripheral region of the phantom. The artifacts are reduced as the energy of the spectrum increases, as expected. Figure~\ref{fig:f6}(f-k) depict the images reconstructed using projection acquired with the six energy bins of the photon counting detector. From energy bin 20-30 keV to 30-40 keV, we can see the attenuation of iodine contrast agent increases strongly, indicating the K-edge property of the iodine, which is at 33.2 keV. From energy bin 40-50 keV to 50-60 keV, the blood insert which is replaced by the gadolinium contrast increases strongly. This is because the K-edge of gadolinium is at 50.2 keV. Note that all images were normalized to Hounsfield units with respect to the value measured in the central water region.
\begin{figure*}[t]
    \centering
    \includegraphics[width=6in]{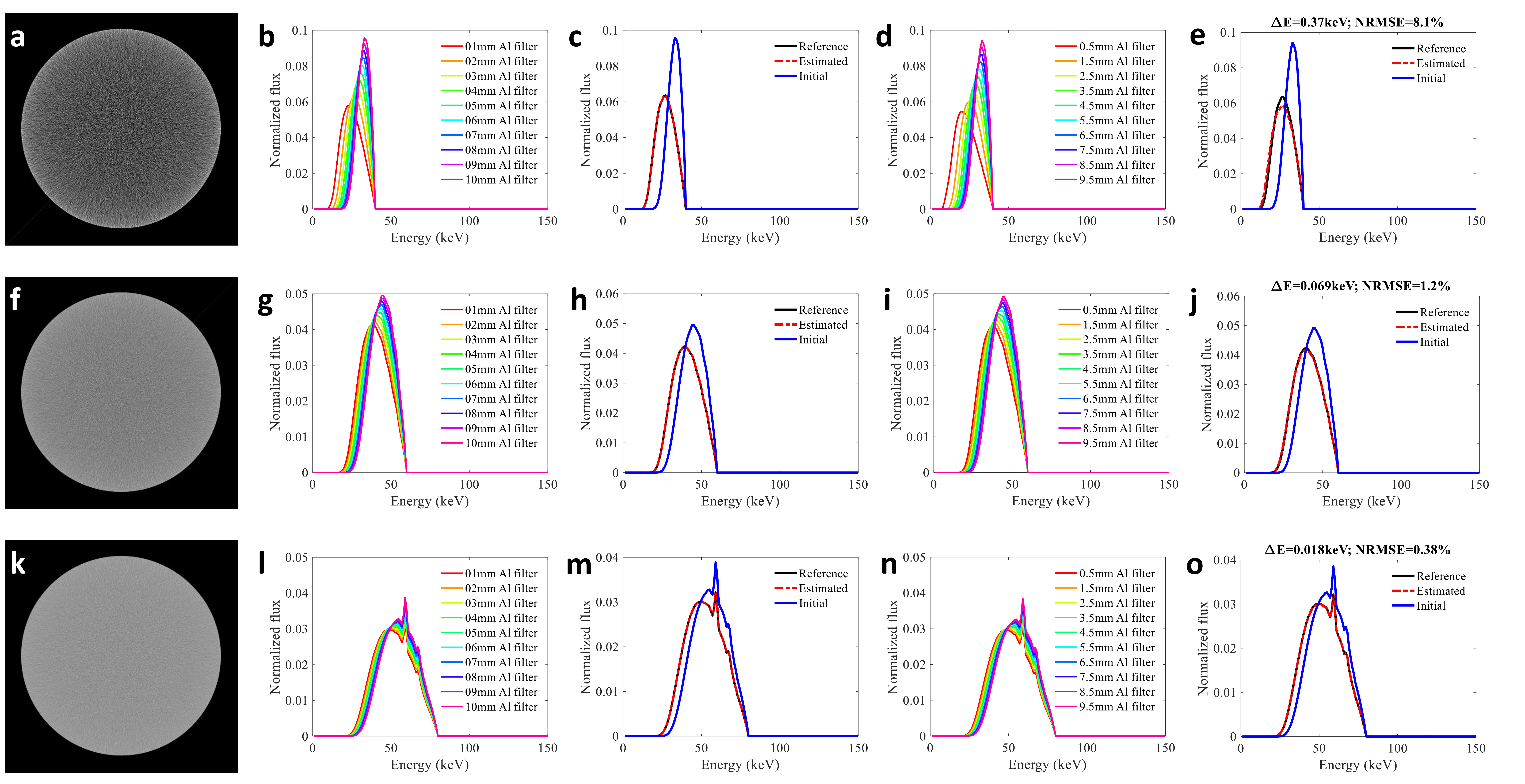}
    \caption{\ihighlight{Energy spectra estimation using a water phantom with different model spectra. The first, second and third rows show spectra estimation for 40, 60, and 80 kV, respectively. (\textbf{c}, \textbf{h}, \textbf{m}) are spectra estimated using model spectra (\textbf{b}, \textbf{g}, \textbf{l}), respectively. (\textbf{e}, \textbf{j}, \textbf{o}) are spectra estimated using model spectra (\textbf{d}, \textbf{i}, \textbf{n}), respectively.}}
    \label{fig:spek}
\end{figure*}

\begin{figure*}[t]
    \centering
    \includegraphics[width=\textwidth]{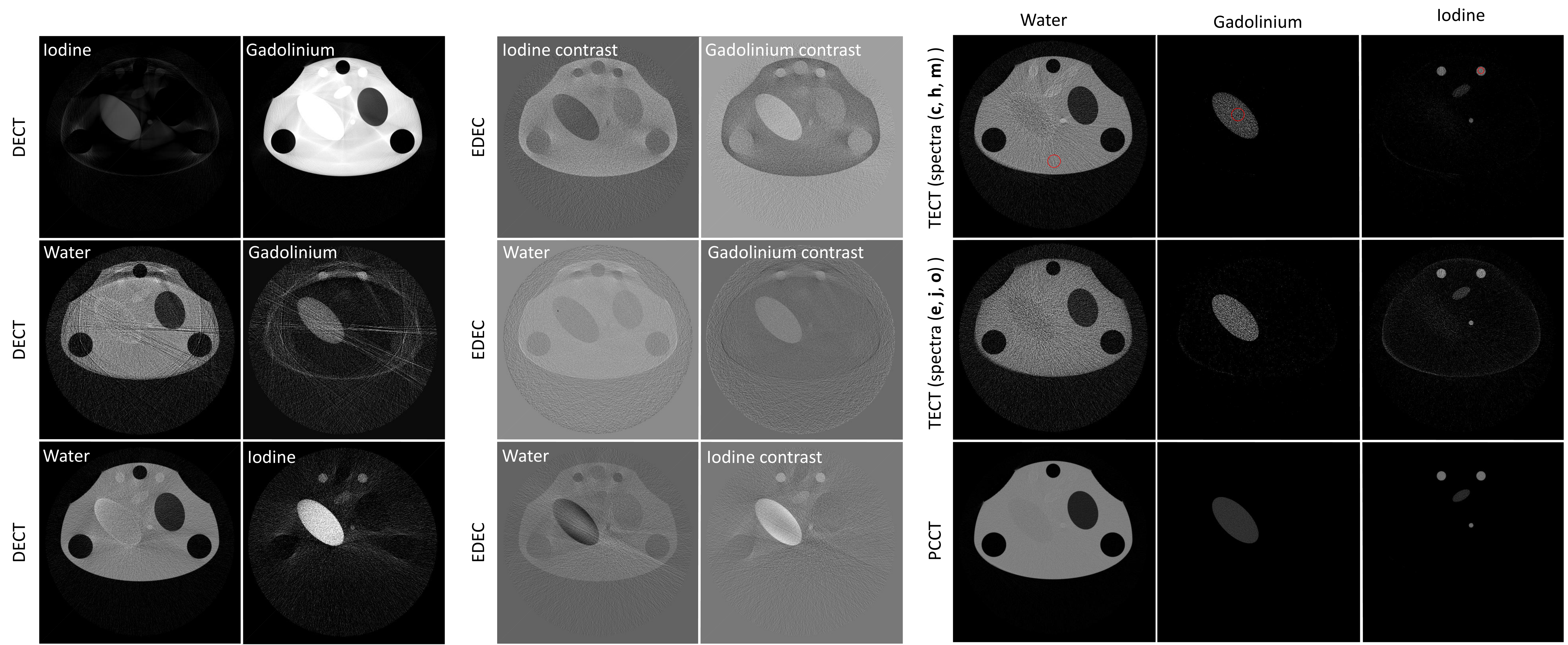}
    \caption{\ihighlight{Results of multi-contrast material decomposition using DECT, \icomment{EDEC}, TECT and idealized PCCT data. DECT and \icomment{EDEC} cannot obtain reasonable material-specific images, while both TECT and PCCT can provide accurate material-specific images. For DECT, TECT and PCCT, Water, gadolinium and iodine images are windowed to [0, 2], [0.05, 0.15], and [0.002, 0.02] g/cm$^{3}$, respectively.}}
    \label{fig:f7}
\end{figure*}

\ihighlight{Figure~\ref{fig:spek} shows triple-energy spectra estimated using a uniform water cylinder with two sets of model spectra. The first, second and third rows show spectra estimation for 40, 60 and 80 kV, respectively. Figure~\ref{fig:spek}(c, h, m) are spectra estimated using model spectra Fig.~\ref{fig:spek}(b, g, l). As can be seen, the estimated spectra match with the reference spectra quite well. Figure~\ref{fig:spek}(e, j, o) are spectra estimated using model spectra Fig.~\ref{fig:spek}(d, i, n). In this case, the model spectra do not contain the reference spectrum, which is much more difficult to accurately recover the reference spectra. Hence, the estimated spectra in Fig.~\ref{fig:spek}(e, j, o) are not as accurate as spectra in Fig.~\ref{fig:spek}(c, h, m). Mean energy differences $\Delta E$ between the estimated spectra and reference spectra are 0.37, 0.07 and 0.02 keV for the 40, 60 and 80 kV spectra, respectively. Normalized root mean square errors between the estimated spectra and the reference spectra are 8.1\%, 1.2\%, and 0.4\%, respectively. }

\ihighlight{Figure~\ref{fig:f7} shows the results of material decomposition using DECT, TECT and idealized PCCT data. As can be seen, for DECT and \icomment{EDEC} material decomposition, no matter how to choose the basis materials (iodine/gadolinium, water/gadolinium and water/iodine), one cannot obtain reasonable material-specific images. For TECT and PCCT material decomposition, water, gadodiamide and pure iodine are selected as the basis materials. Both methods can provide superior material-selective images. In addition, TECT material decomposition using spectra estimated with different model spectra show comparable image quality, suggesting the method is robust with respect to spectrum estimation. Quantitative measurements indicate TECT decomposed densities of the materials are comparable with that obtained by PCCT, as shown in Table~\ref{tab:mousephantom}. }%For the gadolinium and iodine contrast agents, the measured densities are 69 mg/mL and 10 mg/mL for PCCT, and 72 mg/mL and 11 mg/mL for TECT, respectively. For material images decomposed using TECT with spectra shown in Fig.~\ref{fig:spek}(e, j, o), the gadolinium and iodine concentrations are 78 mg/mL and 11 mg/mL for PCCT, respectively.} %

\begin{table}
\vspace{0em}
\caption{Quantitative measurements of the material-specific images of the mouse phantom decomposed using DECT, TECT and PCCT. TECT$^1$, TECT$^2$ correspond triple-energy material decomposition using spectra show in Fig.~\ref{fig:spek}(c, h, m) and (e, j, o), respectively.}
\label{tab:mousephantom}
\begin{center}
\begin{tabular}{c|c|c|c} \toprule
\hline
\multicolumn{2}{c|}{Density and concentration}  & Measured & Difference \\ \hline

\multirow{4}{*}{Water (g/cm$^3$)}  & DECT  & 1.00 & 0 \\
                        & TECT$^1$  & 0.98 & -0.02 \\
                        & TECT$^2$  & 0.89 & -0.11 \\
                        & PCCT  & 1.00 & 0 \\ \hline
 \multirow{4}{*}{ Gadodiamide (mg/mL)}  & DECT  & - & - \\
                        & TECT$^1$  & 63 & 3 \\
                        & TECT$^2$  & 68 & 8 \\
                        & PCCT  & 60 & 0 \\ \hline
 \multirow{4}{*}{Iodine (mg/mL)}  & DECT  & - & - \\
                        & TECT$^1$  & 9 & -1 \\
                        & TECT$^2$  & 11 & 1 \\
                        & PCCT  & 10 & 0 \\ \hline

%\bottomrule
\end{tabular}
\end{center}
\end{table}

\subsection{TwinBeam Dual-source configuration}

\begin{figure}[t]
    \centering
    \includegraphics[width=3.5in]{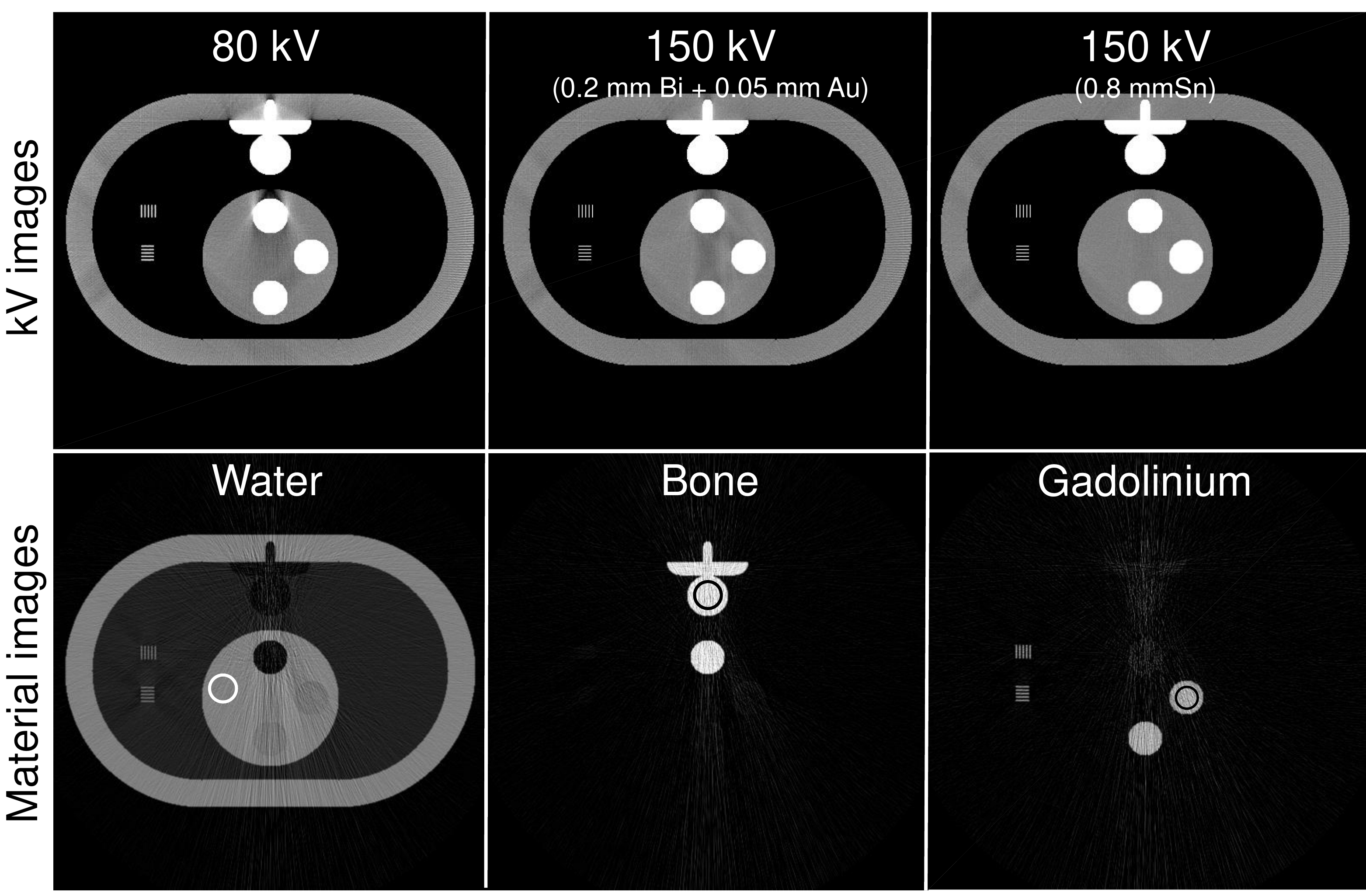}
    \caption{Results of the simplified thorax phantom using triple-energy CT with TwinBeam Dual-source configuration. The labeled circles are used for quantitative measurements. The decomposed densities match with the true densities quite well. All of the kV images are windowed to [-200, 200] HU. Water and bone images are windowed to [0, 2] g/cm$^{3}$, and gadolinium images are windowed to [0, 0.1] g/cm$^{3}$.}
    \label{fig:f8}
\end{figure}

Figure~\ref{fig:f8} shows the results of material decomposition images using triple-energy measurements with "TwinBeam" dual-source configuration. The first row from left to right shows the polychromatic CT images at 80 kV, 150 kV with 0.2 mm bismuth plus 0.05 mm gold filter and 150 kV with 0.5 mm tin filter, respectively. Beam hardening artifacts in the CT images are reduced gradually as the mean energies of the spectra increase. The second row depicts the decomposed images with water, bone, and gadodiamide as basis materials. It is visible that gadodiamide is well decomposed from the contrast agent solution. Quantitative measurements are shown in Table~\ref{tab:thoraxphantom}.%show the relative error of the densities of the basis materials are 0.5\%, 6.0\% and 6.6\% for water, bone and gadodiamide, respectively.

\subsection{Fast kV switching configuration}

\begin{figure}[t]
    \centering
    \includegraphics[width=3.5in]{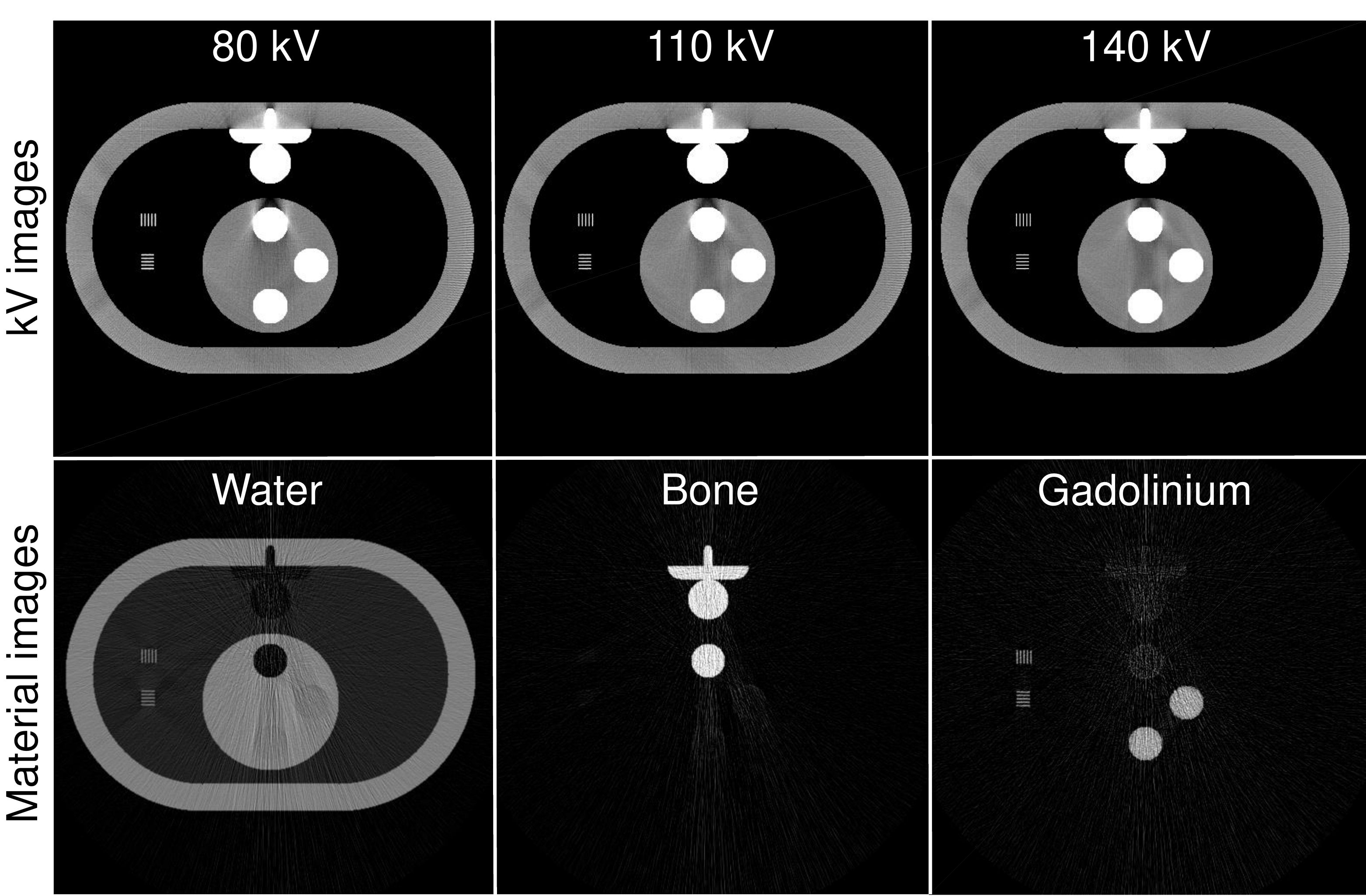}
    \caption{Results of the simplified thorax phantom using triple-energy CT with fast kV switching configuration. The labeled circles are used for quantitative measurements. The decomposed densities match with the true densities quite well. All of the kV images are windowed to [-200, 200] HU. Water and bone images are windowed to [0, 2] g/cm$^{3}$, and gadolinium images are windowed to [0, 0.1] g/cm$^{3}$.}
    \label{fig:f9}
\end{figure}

The results of the simplified thorax phantom using triple-energy CT with fast kV switching configuration are shown in Fig.~\ref{fig:f9}. The first row depicts the polychromatic CT images at 80, 110, and 140 kV. As can be seen, the beam hardening artifacts are present in all of the kV images and do not reduce as the energy increase. This is because the filtration can not be changed in this configuration; resulting in a broad spectrum for the high kV setting. The second row shows the decomposed images with water, bone, and gadodiamide as the basis materials. Gadodiamide is well decomposed from the contrast agent. Quantitative measurements show the decomposed densities match with the true densities quite well (Table~\ref{tab:thoraxphantom}).

\begin{table}
\vspace{0em}
\caption{Quantitative measurements of the material-specific images of the simplified thorax phantom and the anthropomorphic thorax phantom decomposed using TECT with TwinBeam dual-source (TB-DS), fast kV switching (fKV), and TwinBeam dual-source at the presence of a bow-tie filter (TB-DS-BT).}
\label{tab:thoraxphantom}
\begin{center}
\begin{tabular}{c|c|c|c}
\toprule\hline
\multicolumn{2}{c|}{Density and concentration}  & Measured & Difference \\ \hline

\multirow{3}{*}{Water (g/cm$^3$)}  & TB-DS  & 1.006 & 0.006 \\
                        & fKV  & 1.007 & 0.007\\
                        & TB-DS-BT  & 1.035 & 0.035 \\ \hline
 \multirow{3}{*}{Bone (g/cm$^3$)}  & TB-DS  & 1.727 & -0.123 \\
                        & fKV  & 1.724 & -0.126 \\
                        & TB-DS-BT  & 1.641 & -0.209 \\ \hline
 \multirow{3}{*}{Contrast (mg/mL)}  & TB-DS  & 56 & -4 \\
                        & fKV  & 53 & -7 \\
                        & TB-DS-BT  & 55 & -5 \\ \hline

\bottomrule
\end{tabular}
\end{center}
\end{table}

\subsection{Anthropomorphic thorax phantom with bow-tie filter}

\begin{figure}[t]
    \centering
    \includegraphics[width=3.5in]{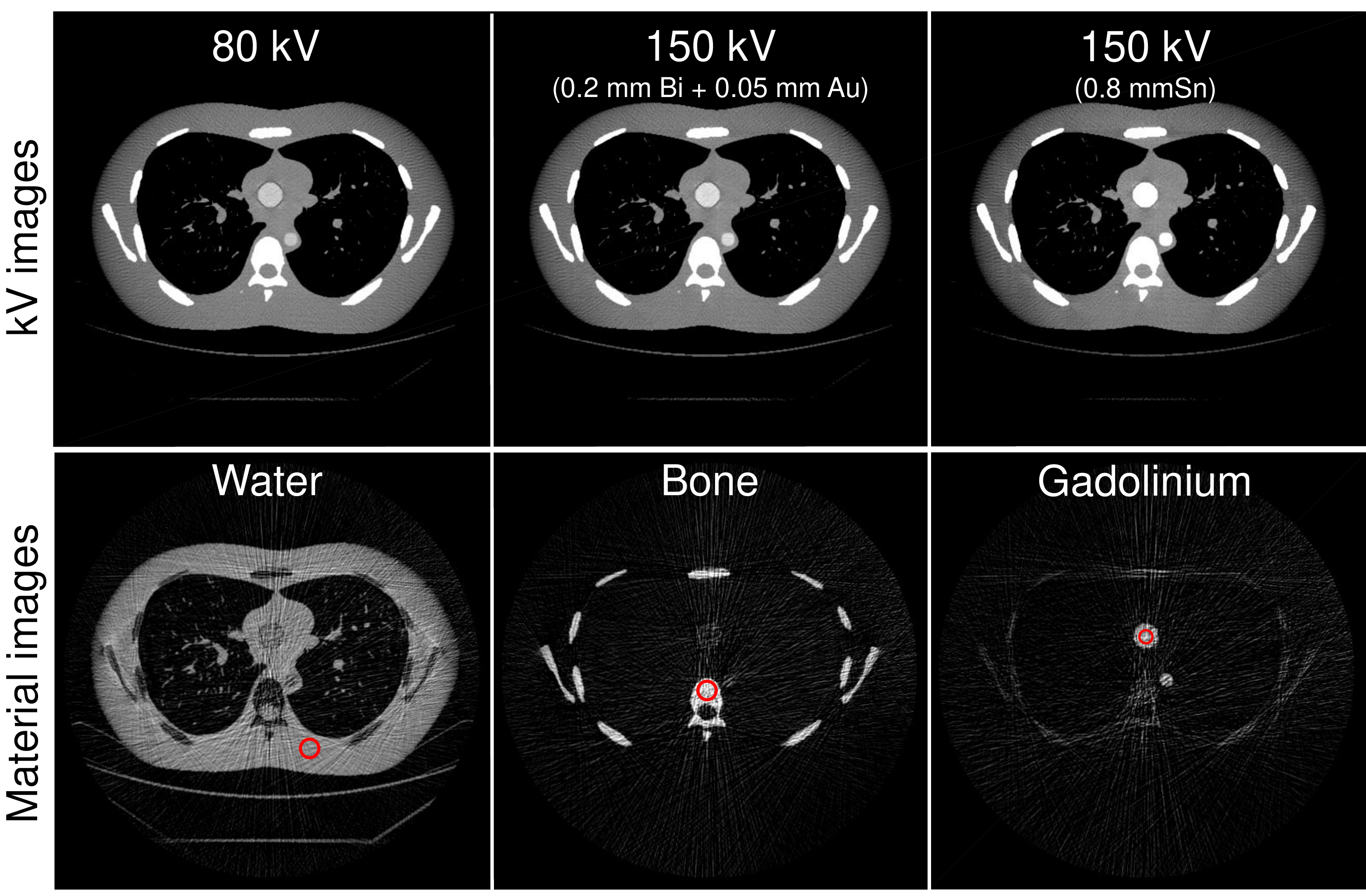}
    \caption{Results of the anthropomorphic thorax phantom using triple-energy CT with TwinBeam Dual-source configuration at the presence of a bow-tie filter. The labeled circles are used for quantitative measurements. The decomposed densities match with the true densities quite well. All of the kV images are windowed to [-500, 500] HU. Water and bone images are windowed to [0, 2] g/cm$^{3}$, and gadolinium images are windowed to [0, 0.1] g/cm$^{3}$.}
    \label{fig:f10}
\end{figure}

Figure~\ref{fig:f10} shows the results of the anthropomorphic thorax phantom using "TwinBeam" dual-source triple-energy configuration at the presence of bow-tie filter. The first row shows the kV images and we can see marginal beam hardening artifacts in all of the images. In addition, the attenuation coefficients (HU value) at the peripheral of the thorax phantom are lower than the other regions which differs from the behavior of the simplified thorax phantom images. This can be attributed to the application of the bow-tie filter, with which the spectrum at the peripheral region is harder than in the central region, resulting in a smaller attenuation contribution in the reconstructed CT images.
%To be more realistic, we further evaluate the proposed method with the application of a bow-tie filter.

The decomposed images in the second row suggest gadodiamide is well separated from the soft tissue and bone. Quantitative measurements show that the decomposed densities are close to the true densities of the basis materials (Table~\ref{tab:thoraxphantom}).

\section{Discussion}

%Discussion about the triple energy implementation, cite ge wang and guohua cao paper (multi-source CT).
%
%Long's multi-material decomposition work joint domain using PL framework

\ihighlight{
Based on currently available dual-source and fast kV switching DECT configurations, we provide a unified framework to perform triple-energy material decomposition, and demonstrated its feasibility using realistic applications with estimated spectra. Multi-contrast imaging shows the benefit of TECT over DECT. DECT and EDEC need a third energy measurement or calibration to provide accurate multi-contrast images. Material images generated by the proposed method are comparable to those generated with the idealized photon-counting detector. With the limitations of current energy-resolved photon-counting detector, the proposed method represents a major step toward realistic TECT imaging and is potential useful for clinical applications.}

%The quantitative virtual non-contrast image and contrast agent uptake provide useful information for clinical applications.
%One would expect that the nonlinear projection-domain method outperforms the image-domain methods on artifacts reduction and contrast-to-noise ratio (CNR)~\cite{kuchenbecker2015}. Hence, both these two matrixes were compared between the image-domain method and proposed projection domain method.

Based on~(\ref{equ:polyreprojBimg}) and~(\ref{equ:calibration}), in order to perform material decomposition as accurate as possible, one has to accurately model the energy spectrum $\Omega(E)$. There are several different ways to obtain the spectrum $\Omega(E)$. The first way is to directly measure the spectrum using energy-resolved photon-counting detectors (such as cadmium zinc telluride and cadmium telluride detectors). However, the accuracy of direct measurement usually suffers from spectral distortion which is caused by charge sharing, pulse pileup, and K-escape energy loss. Hence, it is desirable to model~\cite{lee2016} or correct~\cite{touch2016} the spectrum distortion before material decomposition. Alternatively, one can use direct or indirect transmission~\cite{zhao2015itm} measurements to obtain an effective energy spectrum which incorporates the contribution of inherent filtration and detector response functions. %~\cite{sidky2005,zhang2007,duan2011}

In realistic raw projection data sets, the acquisition with the "TwinBeam" dual-source configuration is inconsistent, which means that the obtained dual- and triple-energy measurements are not from the same X-ray path. However, the proposed method is performed in projection domain which requires consistent raw projection data. To solve this, one may want to combine the proposed method with a material decomposition from inconsistent rays (MDIR) algorithm~\cite{maass2011} which employs an iterative polychromatic forward projection mechanism. It has to be noted that except the available commercial DECT configurations, triple-energy measurements could also be acquired by a multi-source CT geometry which was originally proposed for ultrafast CT imaging~\cite{neculaes2014}.
%inconsistency rays for Siemens projection data. using forward projection

%It is not necessary to employ energy-discriminating detectors to perform K-edge material involved multi-material decomposition.
%
%Compared to the image domain methods or postreconstruction techniques, the proposed projection domain has potential to provide more accurate results,

%Compared to the existing projection-domain or image-domain methods where tedious and time consuming calibration measurements need to be performed time to time, with the incorporation of X-ray energy spectrum, the proposed method does not need dedicated phantom calibration to predetermine the decomposition coefficients for specific material and scan protocol.

In this study, we have used the multi-variable downhill simplex numerical optimization algorithm to solve~(\ref{equ:calibration}). Note that this is a typical least square problem and other optimization algorithms can be applied to solve the problem, such as Levenberg-Marquardt algorithm. For multi-material decomposition, volume conservation~\cite{mendoncca2014} and mass conservation~\cite{liu2009} can be used to regularize the least-square problem, which may be helpful for the optimization procedure. Image domain material decomposition methods are widely used in routine application today. In essence, these methods are based on linear combinations of the reconstructed high- and low-energy CT images. As a comparison, the proposed method provides a nonlinear way to obtain material- and energy-selective images, hence, it may outperform the routinely used linear image-domain methods on contrast-to-noise ratio (CNR) and artifacts reduction~\cite{kuchenbecker2015}. \icomment{Since material decomposition is usually an ill-posed inverse problem, noise is magnified in this procedure, which introduces streaks in the reconstructed material images. Future study will focus on mitigating the streaks by introducing statistical iterative reconstruction and regularized reconstruction~\cite{Sullivan2004, chen2016,ducros2017}. }

%Additionally, by expressing the polychromatic projection using the integral of material images (equation~(\ref{equ:polyreprojBimg}))and further the polynomials (equation~(\ref{equ:polyreprojBimg2})), the least-square formulation provides a framework for material decomposition, including dual- and multi-material decomposition. Thus, if triple-energy measurements are available, the framework can be employed directly for multi-material decomposition. In this case, one only needs to represent the measurements as a vector,
%\beq\label{equ:tripeEnergy}
%\vec{p}_{m}=
%\begin{pmatrix}
%  \vec{p}_{mH}\\
%  \vec{p}_{mM}\\
%  \vec{p}_{mL}
% \end{pmatrix},
%\eeq
%with $\vec{p}_{mM}$ the median-energy attenuation measurements.

%Based on its ability of extension to multi-material decomposition, one may perform material decomposition with elements as the basis material. This may be potential interesting for a variety of applications, including non-destructive testing, explosive material testing and clinical applications. In this case, conservation criteria should be useful to improve the ill-conditioning of the optimization problem. In the future, studies on how to introduce the conservation criteria for the optimization problem will be performed.

In~(\ref{equ:polyreprojBimg}), projection estimation based on the integrals of the material-selective images does not take scatter radiation into account. Hence, the presence of scatter radiation may influence the decomposition procedure, yielding inaccurate line integrals and further inaccurate material-selective images. During the numerical evaluation, scatter radiation is not considered. For realistic applications, one may want to perform scatter correction in front of the material decomposition. To perform scatter correction, the scatter radiation needs to be estimated and then subtracted from the raw projection data~\cite{zhu2009a,zhao2015,zhao2016,shi2017x}. This would be performed for both dual- and triple-energy measurements.

\section{Conclusion}

A nonlinear material decomposition framework for both dual- and triple-energy CT is developed and its feasibility of quantitative material decomposition with estimated spectra is demonstrated using scanning protocols for clinically realistic applications. With energy-integrating detector-based DECT configurations, the method shows robustness against spectrum estimation and can provide material-specific images that are comparable to photon-counting CT. It is promising that many CT-based \ihighlight{diagnostic} and therapy applications will greatly benefit from the proposed material decomposition technique.

%We have developed a nonlinear material decomposition framework for both dual- and triple-energy CT. In this framework, polychromatic projections are estimated with line integrals through material images. The estimated projections are then compared to the measured projections. In this way, the material decomposition problem is formulated as a least-square problem which minimizes the quadratic error between the measured projection and the estimated projection, with the line integrals of the basis material images as the only unknowns. The problem is solved iteratively by updating the weights to minimize the quadratic error. We have evaluated the proposed method using various phantoms and scanning protocols for clinically realistic configurations. The results show that the method can provide quantitative material- and energy-selective images. It is possible to use currently available commercial DECT configurations to perform triple-energy measurements. It is promising that many CT-based \ihighlight{diagnostic} and therapy applications will greatly benefit from the proposed material decomposition technique.

%Hence, the proposed work is feasible for multi-contrast CT imaging and may greatly benefit clinical applications

% if have a single appendix:
\section{Appendix}
\label{sec:appen}
\icomment{The optimization strategy used in this study is Nelder-Mead method which is a numerical method to find the minimum of (4) in a multidimensional space. For two-variable case (DECT), the method initializes with three randomized vertices in the parameters space (note that each vertice has two components). These three vertices form a triangle. The algorithm then compares function values at the three vertices (the best $\textbf{\emph{B}}$, the good $\textbf{\emph{G}}$, and the worst $\textbf{\emph{W}}$) of the triangle, i.e., $f(\textbf{\emph{B}})<f(\textbf{\emph{G}})<f(\textbf{\emph{W}})$. The worst vertex, where the objective function value is maximum, is rejected and replaced with a new vertex. The operations to generate a new vertex include: reflect, extend, contract and shrink. All of the operations are based on the current triangle. With the new vertex, a new triangle is formed, and the search is continued. The process generates a sequence of triangles, for which the function values at their corresponding vertices get smaller and smaller, until the minimum vertex (the optimal $A_i$) are found.} The algorithm used to solve (\ref{equ:calibration}) can be summarized as follows:
\begin{adjustwidth}{0.6cm}{}
\textbf{Algorithm}\\
1. Initialize $A_i^{(0)}$;\\
2. Randomly initialize three points (DECT) in the parameter space for the first triangle: $A_i^{(0)}[k],~k=1,2,3$; \\
3. Calculate objective function values: $f(A_i[k])\leftarrow A_i^{(0)}[k]$;\\
4. Ranking:  $f(\textbf{\emph{B}})<f(\textbf{\emph{G}})<f(\textbf{\emph{W}})\leftarrow f(A_i[k])$;\\
5. Calculate the reflect point $f(\textbf{\emph{R}})$;\\
6. \textbf{WHILE} (criterion is satisfied)\{\\
7. ~~~IF($f(\textbf{\emph{R}})<f(\textbf{\emph{G}})$)\{\\
8. ~~~~~~IF($f(\textbf{\emph{B}})<f(\textbf{\emph{R}})$)\\
9. ~~~~~~~~~~replace \textbf{\emph{W}} with \textbf{\emph{R}}\\
10.~~~~~~ELSE\{\\
11.~~~~~~~~~~compute extend point \textbf{\emph{E}} and $f(\textbf{\emph{E}})$\\
12.~~~~~~~~~~IF($f(\textbf{\emph{E}})<f(\textbf{\emph{B}})$)\\
13.~~~~~~~~~~~~~~replace \textbf{\emph{W}} with \textbf{\emph{E}}\\
14.~~~~~~~~~~ELSE\\
15.~~~~~~~~~~~~~~replace \textbf{\emph{W}} with \textbf{\emph{R}}\}\\
16.~~~\}\\
17.~~~ELSE\{\\
18.~~~~~~IF($f(\textbf{\emph{R}})<f(\textbf{\emph{W}})$)\\
19.~~~~~~~~~~replace \textbf{\emph{W}} with \textbf{\emph{R}}\\
20.~~~~~~compute center point \textbf{\emph{C}} and $f(\textbf{\emph{C}})$\\
21.~~~~~~IF($f(\textbf{\emph{C}})<f(\textbf{\emph{W}})$)\\
22.~~~~~~~~~~replace \textbf{\emph{W}} with \textbf{\emph{C}}\\
23.~~~~~~ELSE\{\\
24.~~~~~~~~~~compute shrink point \textbf{\emph{S}} and $f(\textbf{\emph{S}})$\\
25.~~~~~~~~~~replace \textbf{\emph{W}} with \textbf{\emph{S}}\\
26.~~~~~~~~~~replace \textbf{\emph{G}} with \textbf{\emph{M}}\} ~~~~~//\textbf{\emph{M}} is the midpoint of the good size, i.e., $\textbf{\emph{M}} = (\textbf{\emph{B}} + \textbf{\emph{G}})/2$ \\
27.~~~\}\\
28.\}\\
\end{adjustwidth}

In this work, we have initialized $A_i^{(0)}[k]=0$ for all of studies. The \textbf{WHILE} loop is stopped when either the maximum iteration 200 is satisfied or the difference between $A_i^{(k-1)}$ and $A_i^{(k)}$ is smaller than $10^{-6}$. It takes about 4600 seconds to decomposition DECT sinograms ($1024\times720$) using a typical PC (Intel Core i7-6700K, RAM 32GB).

% use section* for acknowledgement
\section*{Acknowledgment}
This work was partially supported by NIH (7R01HL111141, 1R01 CA176553 and R01E0116777). The contents of this article are solely the responsibility of the authors and do not necessarily represent the official NIH views. This work was also supported by the Natural Science Foundation of China under Grant No. 61601190. The authors are grateful to our colleague Kai Cheng for proofreading the manuscript.
%\vspace{-3.2mm}

% If in two-column mode, this environment will change to single-column format so that long equations can be displayed.
% Use only when necessary.
%\begin{widetext}
%$$\mbox{put long equation here}$$
%\end{widetext}

% Figures should be put into the text as floats.
% Use the graphics or graphicx packages (distributed with LaTeX2e).
% See the LaTeX Graphics Companion by Michel Goosens, Sebastian Rahtz, and Frank Mittelbach for examples.
%
% Here is an example of the general form of a figure:
% Fill in the caption in the braces of the \caption{} command.
% Put the label that you will use with \ref{} command in the braces of the \label{} command.
%
% \begin{figure}
% \includegraphics{}%
% \caption{\label{}}%
% \end{figure}

% Tables may be be put in the text as floats.
% Here is an example of the general form of a table:
% Fill in the caption in the braces of the \caption{} command. Put the label
% that you will use with \ref{} command in the braces of the \label{} command.
% Insert the column specifiers (l, r, c, d, etc.) in the empty braces of the
% \begin{tabular}{} command.
%
% \begin{table}
% \caption{\label{} }
% \begin{tabular}{}
% \end{tabular}
% \end{table}

% If you have acknowledgments, this puts in the proper section head.
%\begin{acknowledgments}
% Put your acknowledgments here.
%\end{acknowledgments}

% Create the reference section using BibTeX:
%\bibliography{dualenergyCT}

\end{document}